\documentclass[draft]{agujournal2019}
\usepackage{url} 
\usepackage[inline]{trackchanges} 
\usepackage{soul}
\usepackage{tikz}
 
\usepackage{textcomp}

\usepackage{comment}

\usepackage{graphicx}
\usepackage{mwe}

\draftfalse

\journalname{arXiv.org}

\begin{document}

\title{A machine learning model of Arctic sea ice motions}

\authors{Jun Zhai$^1$ and Cecilia M. Bitz$^1$}

\affiliation{1}{Department of Atmospheric Science, University of Washington}
%\affiliation{2}{Department of Atmospheric Science, University of Washington}

\correspondingauthor{}{bitz@uw.edu}

\begin{keypoints}
\item Arctic sea ice motions can be modeled using a CNN with predictors of previous-day ice velocity, concentration and present-day surface wind.
\item The superiority of CNN over baseline models suggests the importance of non-local connections compared to local point-wise interactions.
\item The success of the CNN model of ice motion suggests potential for combining machine learning with physics-based models to simulate sea ice.
\end{keypoints}

\begin{abstract}
Sea ice motions play an important role in the polar climate system by transporting pollutants, heat, water and salt as well as changing the ice cover. Numerous physics-based models have been constructed to represent the sea ice dynamical interaction with the atmosphere and ocean. In this study, we propose a new data-driven deep-learning approach that utilizes a convolutional neural network (CNN) to model how Arctic sea ice moves in response to surface winds given its initial ice velocity and concentration a day earlier. Results show that CNN computes the sea ice response with a correlation of 0.82 on average with respect to reality, which surpasses a set of local point-wise predictions and a leading thermodynamic-dynamical model, CICE5. The superior predictive skill of CNN suggests the important role played by the connective patterns of the predictors of the sea ice motion.
\end{abstract}

\section*{Plain Language Summary}
Sea ice, the frozen seawater that floats on the ocean, grows in each hemisphere's winter and retreats in the summer but does not disappear in the current climate. The sea ice coverage is discontinuous and thin enough to move with the winds and currents. These movements alter the sea ice cover and transport pollutants, heat, water and salt. Previous studies have advanced mathematical models that represent the physics of how sea ice moves in response to surface winds. In this study, we propose a different type of model, called a convolutional neural network (CNN), which is constructed purely from observational data without explicitly accounting for the underlying physical insights. Unlike other conventional data-driven models that are trained on each geographic point independently, CNN takes into account how a given location connects to its neighbors via patterns. The superior performance of CNN suggests that local ice motions depend on a large-scale pattern of surrounding winds and that data-based methods are a promising alternative to physics-based models. 

\section{Introduction}
As an essential element in polar climate and dynamics, the movement of sea ice distributes pollutants, transports heat, water and salt and affects the polar energy budget by modifying the ice cover. The factors that determine sea ice motion include the sea ice inertia, atmospheric and oceanic stresses, the Coriolis force, the sea surface tilt and the internal ice force that arises from floe interactions such as collisions, rafting and deformation (\citeA{lepparanta2011drift}, \citeA{feltham2008sea}). 

Several studies have sought equations-based models of varying complexity that utilize a wind-ice relation to model sea ice motion. Based on past estimates that sea ice moves with a speed of 2\% of the surface wind speed and at an angle 45\textdegree \ to the right of the surface wind direction in the Northern Hemisphere,
\citeA{thorndike1982sea} proposed a linear relation between the ice velocity and geostrophic winds:
\begin{equation}
\left[\begin{array}{c}
U \\ V\\
\end{array}\right]=A
\left[\begin{array}{cc}
cos\theta & -sin\theta\\
sin\theta & cos\theta\\
\end{array}\right]
\left[\begin{array}{c}
u \\ v\\
\end{array}\right] + 
\left[\begin{array}{c}
\bar{c_u} \\ \bar{c_v}\\
\end{array}\right],
\label{linear}
\end{equation}
where ($U$,$V$) and ($u$,$v$) are the sea ice velocity and geostrophic wind velocity, respectively, for the same day; $A$ and $\theta$ are the scaling factor and turning angle, respectively; and ($\bar{c_u}$, 
$\bar{c_v}$) is the daily mean surface ocean current velocity. On the long-term (multi-month) average, they found that about half of the variance of the ice motion recorded by buoys is directly related to the surface geostrophic wind, while the other half is due to the mean ocean circulation; on shorter time scales (days to months), geostrophic winds alone could explain more than 70\% of the variance of sea ice motion in their study region. At the daily time scale, the unexplained variance is even less than 20\%. Within 400 km of the coasts, however, the ice inertia or coastal stress gradients can be as important as the wind stress, which disqualifies the wind-only approximation.

Compared to daily sea ice velocities recovered from passive microwave satellite images, \citeA{kimura2000relationship} showed that Eq. \ref{linear} generally explains 70\% to 90\% of the sea ice velocity variance over the Arctic, except along some coastal regions. They found that the spatial variation of $A$ depends on the internal ice stress gradient, which depends on ice thickness and concentration. Generally speaking, thicker and higher sea ice concentration result in a greater internal ice stress gradient, leading to a smaller $A$. For example, according to \cite{kimura2000relationship}, sea ice moves at 2\% of the surface wind speed over the seasonal ice zones but at less than 0.8\% over the Arctic interior. 

Similar conclusions were drawn for Antarctic sea ice from other studies. For example, in the Weddell Sea, nearly 70\% to 95\% of the variance of daily ice drift velocity can be linearly related to the wind velocity, except when the wind speed is below 3.5 m/s \cite{kottmeier1992wind}. In addition, statistically significant relations between the sea ice velocity and local winds were detected from observations in most sectors over the Antarctic \cite{holland2012wind}. 

The wind-ice motion relation has been reexamined for the Arctic in different contexts by later studies (e.g. \citeA{lepparanta2011drift}, \citeA{spreen2011trends}, \citeA{kwok2013arctic}). \citeA{park2016analytical} introduced a more complex nonlinear analytical model that calculates the ice velocity given that the surface winds are known and the ocean geostrophic currents are weak. 

In contrast to the algebraic models described so far, another class of physics-based models computes the sea ice velocity from a differential equation, namely the ice momentum equation, which is primarily driven by surface stresses from winds and ocean currents (e.g., see Hibler, 1979). \nocite{hibler1979} Specifically,  
\begin{equation}
m \frac{\partial\mathbf{u}}{\partial t}=\nabla \cdot \vec{\sigma}+\vec{\tau_a}+
\vec{\tau_o}-\hat{k}\times mf \mathbf{u}-mg \nabla H_o,
\label{momentum_equ}
\end{equation}
where $\vec{\sigma}$, $\vec{\tau_a}$ and $\vec{\tau_o}$ are the internal stress tensor, atmosphere stress, and ocean stress, respectively; $m$ is the combined mass of ice and snow per unit area; $H_o$ is the sea surface height; and $\mathbf{u}$ represents (u, v). The last two terms on the right hand side are the Coriolis force and sea surface tilt term. %The parameterization for the atmosphere and ocean stresses contains the sea ice concentration as a multiplicative factor in order to accommodate the free drift scenario in low ice concentration regions.
Though having achieved a certain amount of success, models utilizing the momentum equation face challenges in practice, which include, but are not limited to, justification for assumptions about the sea ice rheology (via the constitutive law) and the requirement for high-resolution observations to verify the large set of parametrizations (\citeA{notz2012challenges}). 

In this study, our goal is to model sea ice motion with an alternative approach that is data-driven and free of physical assumptions. We hypothesize that the daily Arctic sea ice motion is somewhat persistent with substantial influence from the current surface wind, and we seek to determine whether recent concentration and thickness are also key factors. Thus, similar to other models reviewed in this Introduction, we aim to construct a model that ``predicts" sea ice at the present time from a set of available predictors. Here we present our work to predict the instrumental record. Eventually we envision pairing our sea ice motion model with an Earth system model that can provide predictors in the future, with our model replacing the sea ice momentum and constitutive equations in a hybrid physics and machine learning model of the Earth system.

The sea ice motion model we develop is a convolutional neural network (CNN), which can capture nonlinear spatially heterogeneous relationships. We seek to determine if a CNN model trained on limited observational measurements has the potential to outperform physics-based models. We also seek to learn how CNN compares to baseline estimates from persistence and local point-wise or ``pixel-based" machine learning models. 
The structure of the paper is as follows: Section 2 describes the concept and architecture of the CNN as well as the data used for training and testing; section 3 presents the training results and discussions; and section 4 has conclusions. 

\section{Methodology and Data}
CNN is a class of deep-learning neural network first introduced to classify the handwritten documents in the field of computer vision \cite{lecun1998gradient} and later mostly applied to image analysis. Inspired by the biological process whereby animals visualize images through the connectivity between neurons, CNN breaks-down an image composed of complex patterns into a smaller and simpler pattern of input blocks by employing a mathematical operation called ``convolution". The basic architecture of CNN consists of the following elements:
\begin{itemize}
    \item \textbf{an input layer} that is a tensor with dimensions of (number of image samples)$\times$(image height)$\times$(image width)$\times$(number of input predictors/channels),
    \item \textbf{convolutional layers} that convolve the original images with kernels and activation functions into smaller-sized input maps,  
    \item \textbf{pooling layers} that further reduce the dimensions of the input maps via down sampling, and
    \item \textbf{a fully connected layer} that, via multiple layer perceptrons, fully connects the flattened (i.e., collapsing a multi-dimensional array into one dimension) output tensor from the last convolutional layer to the terminal output of the CNN.
\end{itemize}

Our objective is to predict present-day sea ice velocity ($ui_1$, $vi_1$) as accurately as possible with a parsimonious set of predictors as input to the CNN. After a thorough exploration of predictors (documented in \textit{Supporting Information}), we select the following variables as input:
\begin{itemize}
    \item present-day surface wind velocity ($ua_1$, $va_1$),
    \item previous-day sea ice velocity ($ui_0$, $vi_0$), and
    \item previous-day sea ice concentration ($c_0$).
\end{itemize}

We reconstruct the CNN model based on data sets available from satellite-based observations and atmospheric reanalysis from 1990 to 2018. We use the daily averaged 10-m surface winds from the Japanese 55-year Reanalysis (JRA55), \nocite{kobayashi2015jra} which has a resolution of 47 km on average in the Arctic, National Snow Ice Data Center (NSIDC) version 4 multi-source daily sea ice motions  \cite{tschudi2019polar} and passive microwave sea ice concentrations from the NASA team \cite{cavalieri1996sea}. The ice data are regridded from their 25-km grids to the JRA55 grid using Python xESMF package  (https://doi.org/10.5281/zenodo.1134365). The NSIDC sea ice motion data are considered the "truth" for the purposes of our evaluation.

%----describe architecture----
\begin{figure}
\centering
\noindent\includegraphics[scale=0.45,trim={4cm 0cm 0cm 0cm}]{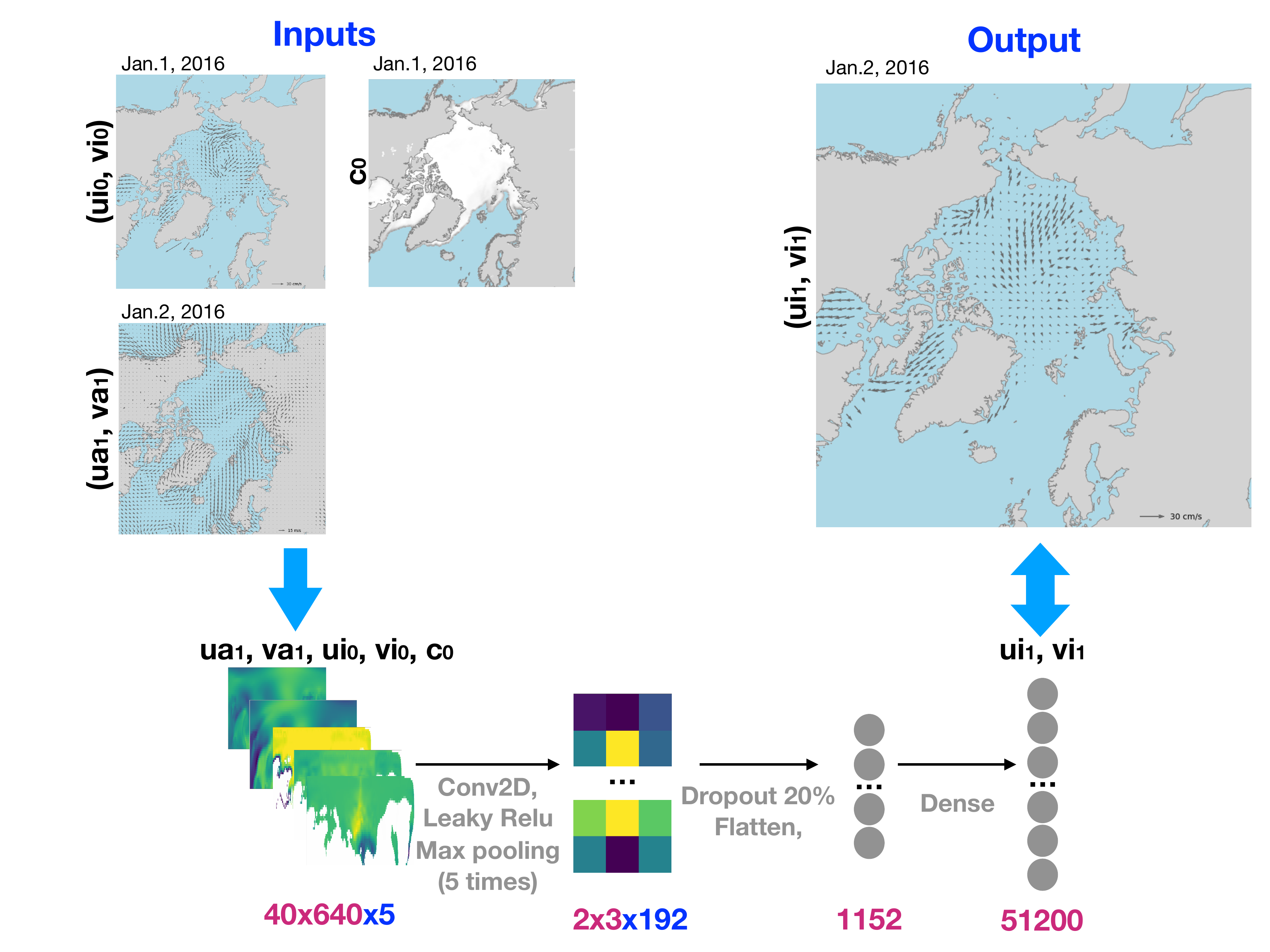}
\caption{The CNN architecture (bottom row). In each layer, the numbers in red are the dimensions of each channel, i.e. (image height)$\times$(image width), and the numbers in blue specify the number of input predictors/channels. An example of inputs and output is presented on the top row. The inputs include the sea ice velocity ($ui_0$, $vi_0$ and $c_0$) on Jan. 1, 2016, while the output is the sea ice velocity ($ua_1$, $va_1$) on Jan. 2, 2016.}
\label{f:architecture}
\end{figure}

We use data that span 1990 to 2018 and split it into a training set (1990 to 2014), a validation set (2015 to 2016) and a testing set (2017 to 2018). The architecture of the CNN is shown in Fig. \ref{f:architecture}: the input is composed of five variables, which are the present-day surface zonal and meridional velocities, the previous-day ice zonal and meridional velocities, and the previous-day ice concentration over the Arctic field of size 40$\times$640. The input is then  processed through five consecutive layers repeating a block unit: a two-dimensional (2D) convolutional layer, a LeakyReLU (Leaky Rectified Linear Unit) layer and a 2D max-pooling layer. The output of this five-time repeated block unit is then passed to a 20\% drop-out layer that  before getting flattened to a one-dimensional (1D) vector with a length of 1152. The final step is to regress the flattened 1D vector to another 1D layer of a size 51200, which is the concatenated 1D vector of two flattened 40$\times$640 images (2$\times$40$\times$640=51200) that represent the present-day zonal and meridional ice velocities, i.e. the target of our prediction. 

% The kernel size for the 2D convolutions and max poolings are (2,3) or (2,2), where the latter is used for later layers which make image size shrink as the convolution progresses, and strides = (1,1) uniformly for all convolutions and max poolings.
The convolutions and max poolings cause the 2D image size to shrink via a kernel size of (2,3) or (2,2) (where the latter is used for later layers) and uniform strides of (1,1). The convolutional layers each use a linear activation followed by a LeakyReLU added as a nonlinear activation with a negative slope coefficient, $\alpha$=0.1. A root mean square error normalized by the standard deviation of the truth (NRMSE, i.e., the second term in Eq. 4) is used as the loss function for optimization with the Adam optimizer. The total number of trainable parameters is 59,132,828. 

Implemented in Python with the Tensorflow/Keras library (https://keras.io), the training takes 50 epochs with a batch size of 365 days on 9862 daily fields (from 1990 to 2016), the last two years of which (2015 to 2016) are used for validation. Then a fresh new set of daily fields (2017 and 2018) are used for final testing. To avoid obtaining accidental results due to fixed test sets, we have also performed a set of randomizations by shuffling the years for training, validation and testing. For example, one of the randomized set uses 1992 to 2016 for training, 2017~2018 for validation, and 1990 to 1991 for testing, etc. The results confirm that the prediction skill is not sensitive to the randomization and maintains stable metric scores. Therefore, here we present one set of training only by focusing on the results obtained using 1990 to 2016 for training and validation and 2017 to 2018 for final testing. 

%---baseline models------
To evaluate the CNN performance, we set-up six baseline models for comparison: persistence (PS), damped persistence (DPS), linear regression (LR), random forest (RF), multiple layer perceptron (MLP) and the 5th version of the Los Alamos Sea Ice Model \cite{Hunke2015} (CICE5). PS predicts the present-day sea ice velocity the same as that of the previous-day. DPS adds a damped anomaly to a 365-day daily climatology of the sea ice motions, with the damping coefficent of a 1-day lag correlation of the anomalies (\citeA{wang2013seasonal}). LR, RF and MLP all regress the present-day sea ice velocity on the five input predictors used by CNN plus sine of latitude and cosine of longitude as an attempt to provide these models some information about location. The fitting of LR, RF and MLP merges input data in time and space as an independent sample. It is worth noting that one advantage of CNN over these baseline models lies in the fact that instead of processing each grid point individually, CNN processes them in blocks that distill the neighboring connections into useful information. 

To make the CICE5 model output in sync with observations and therefore comparable to the CNN model and other baseline models, it is run with prescribed atmospheric reanalysis from JRA55 for the same time period (1990-2018). We tested the effect of replacing the ice motion computed by CICE5 every day at midnight with NSIDC sea ice motions for the previous day, i.e., $(ui_0,vi_0)$, so the ice motion was initialized each day in CICE5 with the NSIDC sea ice motion and evaluated 24 hours later. We found a negligible effect as expected because sea ice motion reaches an approximate steady state in only a few hours after a wind-stress change \cite{campbell1965}; therefore, we did not use this technique in CICE5 for the remainder of this study.  

In summary, the CNN model and each of the baseline models (including the discretization of Eq. 2 in CICE5) can be considered a model that predicts present-day sea ice motion as a function of a set of possible predictors:
$  (ui_1, vi_1) = f(ua_1, va_1, ui_0, vi_0, c_0, ...)$,
where the actual set of predictors may be a subset of those shown (as for PS and DPS) or may includes many others such as the sea ice thickness distribution, roughness, etc. (as for CICE5).

As for skill metrics, the Pearson correlation (Corr),
\begin{equation}
    Corr=\frac{\overline{(x_i-\bar{x})(y_i-\bar{y})}}
    {\sqrt{\overline{(x_i-\bar{x})^2}}\sqrt{\overline{(y_i-\bar{y})^2}}},
\label{eq:Pearson}
\end{equation}
%(Equation \ref{eq:Pearson}) 
and Skill,
\begin{equation}
    Skill=1-\frac{\sqrt{\overline{(y_i-x_i)^2}}}{\sqrt{\overline{(x_i-\bar{x})^2}}},
\label{eq:Skill}
\end{equation}
%(Equation \ref{eq:Skill}) 
are used to evaluate the model. For a given sample size $n$ of a random variable $x$ (i.e., the truth) indexed with $i$ representing time, the skill of its prediction $y$ can be quantified as (1) the covariance between the prediction and the truth scaled by their individual standard deviations (i.e., Corr) and (2) the the percentage of the true standard deviation explained by $y$ (i.e., Skill). Practically, $x$ and $y$ are flattened 1-D representations of $ui_1$ and $vi_1$ stacked one after the other. The `` $\bar{ }$ ''s in Equation (\ref{eq:Pearson}) and (\ref{eq:Skill}) denote the sample mean. $Corr$ has a range from $-1$ to $1$, with $1$ being a perfect linear scaling, $0$ as no relation and $-1$ as an opposite linear scaling with respect to the original signal. $Skill$ is equivalently $1-NRMSE$. It has an upper bound of 1 and no lower bound, with 1 meaning a perfect prediction, 0 indicating an equivalent skill as predicting the mean, and a negative value indicating a worse skill than predicting the mean. The interpretation of the metrics is that Corr quantifies how much $x$ and $y$ vary together regardless of their relative magnitudes, while Skill takes the predicted magnitude into account. Corr and Skill are then averaged over space and/or time to reduce them to a spatial distribution, time series, or single value.

\section{Results}
%----loss function------
\begin{figure}
\centering
\noindent\includegraphics[scale=0.45,trim={0cm 0cm 0cm 2cm}]{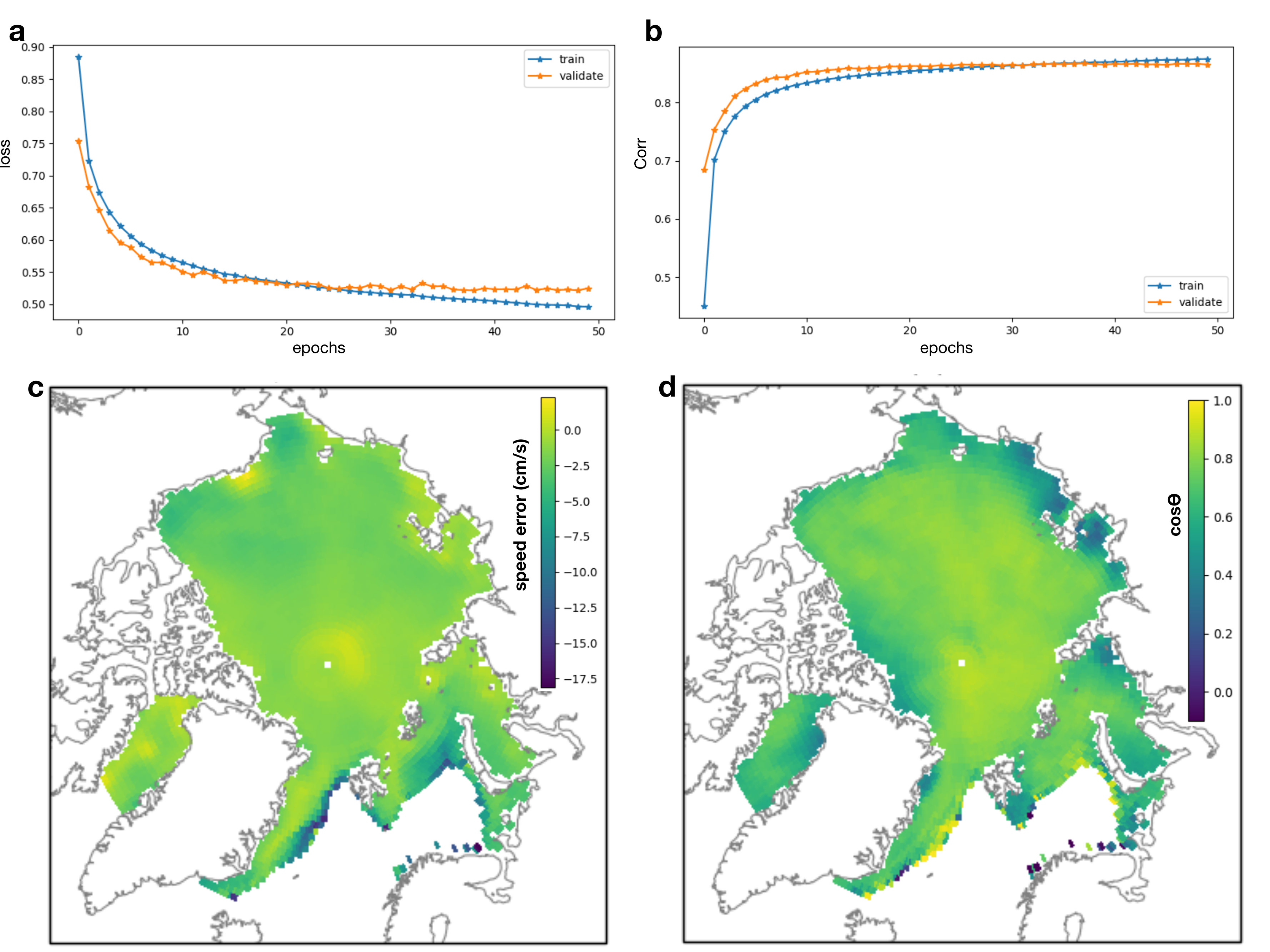}
\caption{The top row shows \textbf{(a)} loss function and \textbf{(b)} Corr of prediction in the training and validation sets as indicated by the legends. The bottom row shows the mean prediction errors in terms of \textbf{(c)} speed and \textbf{(d)} angle averaged over the test data set. Specifically, (c) shows the difference of the predicted speed and the true speed in cm/s with negative values associated with underestimating the speed; (d) shows the cosine of $\theta$, the angle between the predicted velocity and the true velocity, with a $cos\theta$ closer to 1 associated with a smaller angular deviation from the truth and a negative $cos\theta$ associated with an angular deviation of greater than 90\textdegree. Both (c) and (d) are calculated for a given grid point at each time and temporally averaged over the two testing years.}
\label{f:combined}
\end{figure}

We begin by discussing the degree to which the CNN model reaches a stable solution by the end of the 50 epochs that we prescribe.
From the learning curves (Fig. \ref{f:combined}a,b), we see that the minimization of the loss function saturates at around 30 epochs with approximately 50\% of the standard deviation unexplained, which corresponds to Corr of $\sim$0.88 and Skill of $\sim$0.52 for the CNN model output with respect to the truth. The learning curves for training and validation periods are similar to each other after a few epochs, confirming that the model does not suffer from overfitting. In addition, to check the sufficiency of the number of years of used for training, we perform a set of sensitivity experiments to determine the scaling of model skill as a function of the training data size. It follows that the model skill has reached a plateau for a training data size of 25 years (see \textit{Supporting Information} Figure S1), confirming that our training period (1990 to 2014) is sufficient.

%----compare to all baselines (Pattern)---
\begin{figure}
\centering
\noindent\includegraphics[scale=0.35]{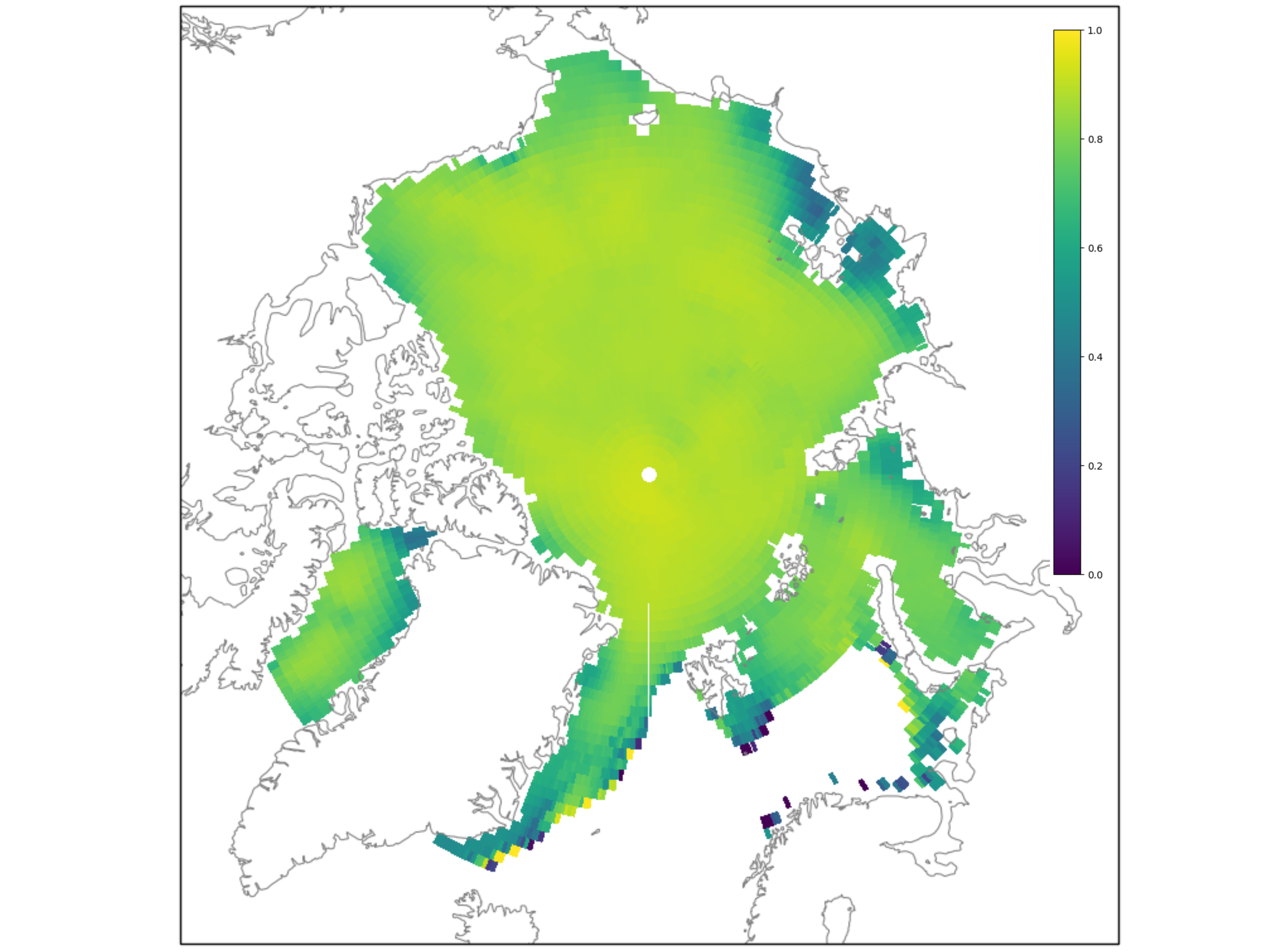}
\caption{Local-wise correlations of the predicted sea ice velocity with respect to the observations. At each grid point, the time series of the predicted sea ice zonal and meridional velocities are concatenated into a 1-D vector and compare to the observation by computing the correlation coefficient. Along the southeast coaste, less than 1\% of the area has negative correlations, and therefore we set the minimum of the colorbar to be 0 for a better visualization.}
\label{f:corrmap}
\end{figure}

\begin{figure}
\centering
\noindent\includegraphics[scale=0.65]{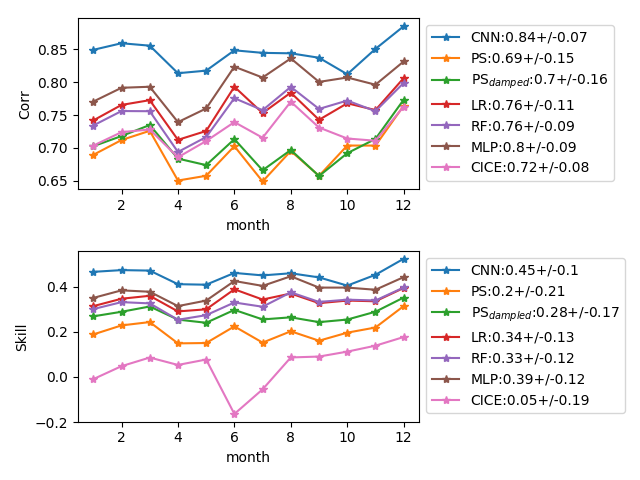}
\caption{The monthly averaged (a) Corr and (b) Skill of daily predictions for different models on the test set. The averages are computed as pattern-wise evaluation for each field given a month. The mean and standard deviation (std) of Corr and Skill over the entire test set for each model are presented in the legends in the form of ``mean+/-std".}
\label{f:compare}
\end{figure}

The predictions are evaluated in two fashions: local-wise and pattern-wise. Local-wise evaluation measures how well the model predicts the time series at a given point and presents a spatial distribution of the prediction metrics, as shown in Fig.  \ref{f:combined}c,d and Fig. \ref{f:corrmap}. Pattern-wise evaluation, on the other hand, quantifies how well the prediction recovers the spatial pattern for each predicted field, quantified by computing Corr or Skill on each field flattened as a 1D vector and averaged over an interval such as a month (Fig. \ref{f:compare}). 

For local-wise evaluation, we evaluate the predicted sea ice velocity in terms of speed and angle. Figure \ref{f:combined}c shows the spatial distribution of the averaged difference between the predicted and true speed. We see that most of the prediction errors of the CNN model manifest underestimating the speed by 5 cm/s or so (approximately 10\% of the average speed). In general, most of the  underestimations occur away from the central Arctic and are greatest (up to 25 cm/s) near the sea ice edge in the Greenland and Barents seas. The significant deviations in the angle prediction, on the other hand, are mostly along the Siberian coast of the Arctic with an angular deviation of up to 100\textdegree \ (i.e. $cos^{-1}(-0.2)\approx100^{\circ}$) from the truth (Fig. \ref{f:combined}d). On average, the angular error by CNN prediction over the interior of the central Arctic is approximately 32\textdegree \ (i.e., $cos^{-1}(0.85)\approx32^{\circ}$). Consistently, Fig. \ref{f:corrmap} shows that the best prediction are made over the central Arctic and greater errors occur near the coasts.

For pattern-wise evaluation, six baseline models (PS, DPS, LR, RF, MLP and CICE5) are used for comparison. As shown in Fig. \ref{f:compare}, CNN outperforms all the baseline models that make local predictions, indicating the advantage gained from CNN's nonlinearity and the neighboring ice and coastline connections. Specifically, the fact that CNN outperforms persistence (PS) confirms the influence of surface winds and ice conditions on the sea ice dynamic response. 

Since a goal of our study is to determine how well a CNN model performs compared to thermodynamic-dynamic sea ice models, we also compute Corr and Skill for the daily prediction of sea ice motion from CICE5 when forced by prescribing the JRA55 reanalysis. We find a Corr of 0.72 and Skill of 0.05 for CICE5 --- about 15\% below Corr and 90\% below Skill of the CNN model. The contrast of high Corr and low Skill suggests a large speed error in CICE5, 4.4 cm/s compared to 3 cm/s on average for the other models (see \textit{Supporting Information} Figure S2). 

A consideration when evaluating the various models is that the data-driven models were both trained and tested on the NSIDC sea ice motions, while the dynamical portion of CICE5 was developed prior to the construction of the NSIDC product. Furthermore, there are errors in the NSIDC sea ice motions owing to errors in the input sources, particularly from satellites and reanalysis winds (see \emph{Supporting Information} Table S2). Ice motions errors from buoy data are thought to be much lower, and NSIDC provides an error estimate of its motion relative to buoy motions. To address this issue of CICE5 being at a disadvantage when evaluated against the NSIDC sea ice motions, we first restricted our computation of Skill and Corr to the locations and times in our test period with the lowest 20\% of errors in the NSIDC sea ice motions. We find that the annual mean skill metrics compared to those in Figure \ref{f:compare} are the same for CICE5 and drops slightly for CNN (from 0.84 to 0.82 for Corr and from 0.45 to 0.37 for Skill). To further explore this issue and motivated by the possibility that errors in the satellite input sources might be higher in summer due to cloudier skies and complications from melt ponds, we retrained the CNN model using data from November to February only. The evaluations of the test years in winter months are very similar to those that use data of all seasons, with CNN outperforming the rest of the models. 

As a final test, we computed the RMSE of sea ice speed for CNN and CICE5 models first for all locations and times and then separately for the upper and lower quantiles of speeds (see \emph{Supporting Information} Table S3 and Figures S2 and S3). The CNN model outperforms CICE5 in all cases, and is particularly successful at lower speeds. 

%Another note is that the low \emph{Skill} obtained by CICE5 is largely due to the mismatch of the speed between the CICE5 prediction and the NSIDC observation, because CICE5 predictions do have a decent \emph{Corr} (top panel of Figure \ref{f:compare}) which evaluate without taking into account the velocity magnitude. In addition, we also provide a few case studies of sea ice motions with low and high speed (see \emph{Supporting Information}.

In summary, given NSIDC sea ice motions as the target for training and evaluations, CNN predicts the sea ice motions in response to the surface winds with the highest stable predictive skills compared to all the other local point-wise baseline models we considered and the state-of-art CICE5 model.

\section{Conclusions}
In this study, we constructed a deep-learning nonlinear model known as a convolutional neural network to predict how the Arctic sea ice responds to a given surface wind. Specifically, it takes the previous-day sea ice velocity and concentration as well as the present-day surface wind as input features and predicts the present-day sea ice velocity. The result of CNN prediction is evaluated against persistence models (PS and DPS), several other fitted models (LR, RF, and MLP), and one differential equation-based model (CICE5). It follows that CNN outperforms all six baseline models in terms of Corr and Skill as the accuracy metrics. 

The superior performance of CNN over the local point-wise models (PS, DPS, LR, RF and MLP) suggests the influence of neighboring sea ice and coastlines across the Arctic when modeling the sea ice motion. Given the NSIDC sea ice motions used as the target, the comparably successful performance by CNN model to the state-of-art CICE5 suggests a potential for combining a machine-learning method (data-driven) with a differential equation-based (physics-based) model to reduce the computational cost. Adjustments to the CNN model presented here based on observational data would be required to match the timestep of the CICE5 model, which is generally one hour. The CNN model's inputs and outputs have a one-day timestep owing to the limited availability of observed sea ice motion and concentration from satellites that were used to train the CNN model. Winds from reanalysis are typically computed sub hourly, though typically reanalysis products are every few hours (e.g., JRA55 is every 3 hours). Interpolating the observed data to a finer time interval may be an option to augment the training data. We verified using synthetic observations with a one-hour timestep that the CNN model is equally successful (documented in \textit{Supporting Information}). Constructing such a hybrid CNN-physics based model is left for future work.

There are a few caveats, however, for CNN. One is that since CNN is a data-driven approach, the success of the model training heavily depends on the availability of the observations or reanalyses of the sea ice motions and the Arctic surface winds, which eliminates applications in regions with measurement deficiency, such as paleoclimate reconstruction. In addition, since CNN is purely data-driven (i.e. fitting with respect to a target ``correct answer"), the accuracy of the training data determines the quality of the prediction. As mentioned in Section 3, there is uncertainty/error associated with NSIDC sea ice motions. It is thus important to understand the error ranges of the training target data (NSIDC sea ice motions) (see \emph{Supporting Information} Table S3). Furthermore, the suboptimal ice motions simulated by CICE5 might relate to tuning conducted in order to account for model variables other than ice motion. When implemented in a complex dynamical system, CNN will inevitably encounter similar compromises in order to be compatible. Thus additional adjustment (e.g. using a loss function that takes into account all aspects of the sea ice system) will be required in order to stabilize the total system outputs. 

%\appendix
%\section{Here is a sample appendix}

\acknowledgments
All data used in this study are publicly available. For climate model output, we use one historical simulation (r1i1p1f1) of the Community Earth System Model version 2 (CESM2, Danabasoglu et al., 2020)\nocite{danabasoglu2020community} in the Coupled Model Intercomparison Project 6 (CMIP6). The 10-m surface winds from the Japanese 55-year Reanalysis (JRA55) are available through \cite{kobayashi2015jra}. Sea ice motion \cite{tschudi2019polar} and passive microwave sea ice concentrations \cite{cavalieri1996sea} are both available from the National Snow and Ice Data Center (NSIDC). CICE5 simulations are uploaded at \cite{zhai_jun_2021_5090804}. We thank the University of Washington College of the Environment for supporting our work.

%Enter acknowledgments, including your data availability statement, here.

\bibliography{references.bib} 

\end{document}